\documentclass[12pt]{article} 
 
\usepackage{epsfig}  
\usepackage{amsmath}
\usepackage{amssymb}

\newcommand{\D}{\displaystyle}

\begin{document} 
 
\begin{titlepage}

\vspace{2cm}

\title
{\vskip -70pt
\begin{flushright}
{\normalsize \ DAMTP--1999--119, {\tt hep-th/9909168}}\\
\end{flushright}
\vskip 20pt
{\bf On the constraints defining BPS monopoles }
}
\vspace{2cm}

\author{
\\
{\large \sc C. J. Houghton}\thanks{e-mail: {\tt C.J.Houghton@damtp.cam.ac.uk}},
\hspace{.2cm}
{\large \sc N. S. Manton}\thanks{e-mail: {\tt N.S.Manton@damtp.cam.ac.uk}}
\hspace{.2cm} {\large and} \hspace{.2cm}
{\large \sc N. M. Rom\~ao}\thanks{e-mail: {\tt N.M.Romao@damtp.cam.ac.uk}}\\
\\
{\normalsize \sl Department of Applied Mathematics and Theoretical Physics}\\
{\normalsize \sl University of Cambridge} \\
{\normalsize \sl Silver Street, Cambridge CB3 9EW, England}
\\
\\}

\date{\large September 21, 1999}

\maketitle
\thispagestyle{empty}
\vspace{1cm}

\begin{abstract}

\noindent
We discuss the explicit formulation of the transcendental constraints 
defining spectral curves of $SU(2)$ BPS monopoles in the twistor
approach of Hitchin, following Ercolani and Sinha. We obtain an
improved version of the Ercolani--Sinha constraints, and show that the
Corrigan--Goddard conditions for constructing monopoles of arbitrary 
charge can be regarded as a special case of these. As an application,
we study the spectral curve of the tetrahedrally symmetric 3-monopole,
an example where the Corrigan--Goddard conditions need to be modified.
A particular 1-cycle on the spectral curve plays an important r\^ole in
our analysis.

\end{abstract}

\end{titlepage}

\section{Introduction}

BPS monopoles for the $SU(2)$ Yang--Mills--Higgs gauge theory have been 
studied for over twenty years, using a number of
different approaches. Twistor methods, relating the solutions of the
integrable differential equations of the model to holomorphic vector
bundles over a so-called twistor space, were first introduced by Ward, 
adapting previous work on the self-duality equations 
for the pure Yang--Mills theory on $\mathbb{R}^{4}$. They enabled
solutions of magnetic charge $k>1$ to be constructed for the first 
time \cite{Wa}.
A twistor approach intrinsic to the geometry of $\mathbb{R}^{3}$ was
developed later by Hitchin in \cite{H1} and~\cite{H2}. In his formulation,
a monopole is associated to a spectral curve, a compact complex curve 
in $T'\mathbb{CP}^{1}$, the total space of the holomorphic tangent bundle of
$\mathbb{CP}^{1}$, satisfying a number of conditions which were
stated in \cite{H2}. 
Based on this approach, new solutions have been
constructed and new characterisations of monopoles developed; we
refer to \cite{Srev} for a brief overview.

The (reduced) moduli space ${\cal N}_{k}$ of gauge-inequivalent BPS
monopoles of a given charge $k$ is a $(4k-1)$-dimensional manifold,
which has been described in several ways. If we adopt the twistor
formulation in terms of spectral curves, it can be characterised 
as the space of complex curves in $T'\mathbb{CP}^{1}$ satisfying a
number of transcendental constraints. For the case where the curve
is nonsingular, Ercolani and Sinha attempted to formulate these 
constraints explicitly in~\cite{ES}; they followed essentially the
method of Hurtubise in~\cite{Hur}, who achieved a satisfactory
description of~${\cal N}_{2}$. Their approach leads to a method of
determining constraints on spectral curves by analysing objects of 
the function theory obtained on them. These constraints parallel the
Corrigan--Goddard conditions \cite{CG} for constructing
$SU(2)$ monopoles, and we shall clarify how they relate to each other.

This paper is organised as follows. 
We start by introducing the relevant aspects of the twistor approach
for monopoles in section~\ref{twistor}, in order to fix the notation. 
In section~\ref{ErcSin}, we review the method of Ercolani and
Sinha, and present a new version (\ref{ourES}) of their constraint 
equations which involves a special 1-cycle $c$ on the spectral 
curve.
The Corrigan--Goddard conditions were originally formulated in
terms of integrals around the equator of $\mathbb{CP}^{1}$, but we
show in section~\ref{CorGod}
how to interpret them equivalently as integrals on the spectral 
curve. Moreover, we establish that the conditions in the 
two methods agree except for one detail: the Corrigan--Goddard
approach enforces $c$ to be of a special sort, namely a combination of
lifts of the equator of $\mathbb{CP}^{1}$ to the spectral curve. 
In section~\ref{4hedral}, we apply the Ercolani--Sinha method to 
compute the spectral curve of the tetrahedral 3-monopole discussed 
in \cite{HMM}.
Thereby, the corresponding 1-cycle $c$ is determined and the result
shows that the Corrigan--Goddard assumption about $c$ is too
restrictive in general; 
we also consider the action of the tetrahedral group 
$\mathfrak{A}_{4} \subset SO(3)$ 
on the homology of the spectral curve and show that it leaves $c$ 
invariant. Finally, we present some concluding remarks in 
section~\ref{cnclsn}.

\section{Twistor methods for BPS monopoles} \label{twistor}

Magnetic BPS monopoles with gauge group $SU(2)$ are defined as gauge
equivalence classes of solutions $(A,\phi)$ to the Bogomol'ny\u\i\
equations in $\mathbb{R}^{3}$
\begin{equation} \label{bgmln}
*F_{A}=\pm \nabla_{A} \phi
\end{equation}
satisfying boundary conditions (see \cite{AH}) that ensure finiteness 
of the energy functional; here $A$ is a connection 1-form (with covariant
derivative $\nabla_{A}$ and curvature 2-form $F_{A}$), and $\phi$ (the
Higgs field) a
function, both taking values in $\mathfrak{su}(2)$. 
Such solutions can be interpreted as particle-like solitons carrying
discrete magnetic charge. They are associated with an integer $k 
\in \mathbb{Z}$ (with $\pm k>0$ according to the sign in equation 
(\ref{bgmln})), which corresponds to the magnetic charge of the
field configuration in suitable units and classifies the solutions 
homotopically; we take $k>0$ throughout. The Bianchi identity together with
(\ref{bgmln}) imply that BPS monopoles are also static classical solutions
of the corresponding Yang--Mills--Higgs theory in the  BPS limit, in 
which the Higgs potential is set to zero, and they correspond exactly
to the minima of the energy functional.

The equations (\ref{bgmln}) are integrable and their solutions can
be studied using methods of complex algebraic geometry. This was 
formulated by Hitchin in \cite{H1} as follows. The space
$\mathbb{T}$ of oriented geodesics (straight lines) of
$\mathbb{R}^{3}$ is a 4-dimensional manifold --- a point on it can be 
specified by a pair of vectors 
$(\mathbf{u},\mathbf{v}) \in \mathbb{R}^{3+3}$, 
where $\mathbf{u}$ has unit
length and defines the orientation of the line, while $\mathbf{v}$
gives the position of the point on the line closest to the origin and
is thus orthogonal to $\mathbf{u}$. This manifold admits a natural
integrable almost complex structure, given at each point 
$(\mathbf{u},\mathbf{v})$
by taking the cross product with $\mathbf{u}$ of each of the 
pair of vectors representing a tangent vector. It turns out that
$\mathbb{T}$, endowed with this complex structure, is isomorphic as
a complex surface to the total space $T'\mathbb{CP}^{1}$ of the
holomorphic tangent bundle to the Riemann sphere. The isomorphism
takes $\mathbf{u}$ to the corresponding point in $\mathbb{CP}^{1}\cong
S^{2}$ and $\mathbf{v}$ to the obvious complex coordinate on the
fibre. We will consider the standard affine pieces $U_{0}$ and $U_{1}$ of 
$\mathbb{CP}^{1}$, identifying the affine coordinate $\zeta$ on $U_{0}$
with the stereographic projection from the south pole and letting $\eta$
denote the corresponding coordinate on the fibre; thus a
tangent vector $\eta  \frac{\partial}{\partial \zeta}| _{\zeta}$ is
assigned the pair $(\eta, \zeta)$. We let $\pi$ be the natural projection 
$\mathbb{T} \rightarrow \mathbb{CP}^{1}$, given in these coordinates 
by $(\eta,\zeta) \mapsto \zeta$, and denote again by
$U_{0}$, $U_{1}$ the pre-images under $\pi$ of the affine pieces of
$\mathbb{CP}^{1}$.

In the literature, $\mathbb{T}$ is often called mini-twistor space.
It admits a real structure $\tau: \mathbb{T} \rightarrow
\mathbb{T}$, which is the anti-holomorphic involution corresponding 
to the reversal of direction of oriented lines in $\mathbb{R}^{3}$;
it obviously has no fixed points.
In terms of our coordinates, it can be seen to be given by
\begin{equation} \label{tau}
\tau: (\eta,\zeta) \mapsto 
(-\frac{\bar{\eta}}{\bar{\zeta}^{2}}, -\frac{1}{\bar{\zeta}}).
\end{equation}
The group $SO(3)$ of rotations in $\mathbb{R}^{3}$ induces an action
on $\mathbb{T}$, which can be easily described in the coordinates
$(\eta, \zeta)$ in terms of the corresponding $PSU(2)$
transformations: The matrix
\[
\begin{bmatrix}
p        &  q \\
-\bar{q} &  \bar{p}
\end{bmatrix}
\in PSU(2), \;\;\;\;\;|p|^{2}+|q|^{2}= 1
\]
acts on the affine coordinate $\zeta$ as
\begin{equation} \label{psu2}
\zeta \mapsto \frac{\bar{p}{\zeta} -\bar{q}}{q \zeta +p}
\end{equation}
and this corresponds to a rotation by $\theta$ around the direction
$\mathbf{n}\in S^{2}$ with 
$n_{1} \sin\frac{\theta}{2}={\rm Im}\,q$,
$n_{2} \sin\frac{\theta}{2}=-{\rm Re}\,q$,
$n_{3} \sin\frac{\theta}{2}=-{\rm Im}\,p$ and
$\cos\frac{\theta}{2}={\rm Re}\,p$; $\eta$ transforms by multiplication
by the derivative of (\ref{psu2}),
\[
\eta \mapsto \frac{\eta}{(q \zeta +p)^{2}},
\]
since it is the fibre coordinate of $T'\mathbb{CP}^{1}$ corresponding
to $\zeta$. 
It is clear from the definitions that the action of
$SO(3)$ commutes with the $\mathbb{Z}_{2}$ action generated by $\tau$.

For each $s\in \mathbb{R}$ and $k\in \mathbb{Z}$, we define a
holomorphic line 
bundle $L^{s}(k)$ on $\mathbb{T}$ through the transition function
\begin{eqnarray}
g^{(s,k)}_{01}: & U_{0} \cap U_{1}  & \longrightarrow \;\; 
\mathbb{C}^{\,*} \nonumber \\
 & (\eta,\zeta) & \longmapsto \;\; 
e^{-s\frac{\eta}{\zeta}} \zeta^{k} \nonumber
\end{eqnarray}
with respect to the trivialising cover $\{U_{0},U_{1}\}$ of
$\mathbb{T}$; this definition is independent of the stereographic
projection used on $\mathbb{CP}^{1}$.
We shall use the notation $L^{s}$ for $L^{s}(0)$ and
${\cal O}(k)$ for $L^{0}(k)$.
These line bundles play a r\^ole in the formulation of the twistor 
correspondence for
monopoles, which we now describe. To a monopole $(A,\phi)$ we
associate the complex vector bundle $E\rightarrow \mathbb{T}$,
whose fibre at an oriented line $\gamma \in \mathbb{T}$ is the 
complex 2-dimensional space of solutions 
$u:\gamma \rightarrow \mathbb{C}^{\,2}$ to the equation
\[
(\nabla_{\gamma}-i\phi)u=0,
\]
where $\nabla_{\gamma}$ is the restriction of $\nabla_{A}$ to
$\gamma$. The Bogomol'ny\u\i\ 
equation (\ref{bgmln}) implies that $E$ is holomorphic; 
it can be regarded as an extension
\begin{equation} \label{ext}
0 \longrightarrow L^{\pm} \longrightarrow E \longrightarrow 
(L^{\pm})^{*} \longrightarrow 0
\end{equation}
of the line subbundles $L^{\pm}\subset E$ of solutions decaying exponentially
as $t\rightarrow \pm \infty$, where $t\in\mathbb{R}$ is the natural
coordinate on $\gamma$. It can be shown that, for any monopole
of charge $k$, $L^{\pm}$ is isomorphic to $L^{\pm 1}(-k)$;
different monopoles correspond to different extensions $E$. 
Given the two short exact sequences (\ref{ext}), we consider the
composite morphism $L^{-}\rightarrow E \rightarrow (L^{+})^{*}$, which 
defines a holomorphic section $P$ of the line bundle 
$(L^{-} \otimes L^{+})^{*} \cong {\cal O}(2k)$. It will determine a
compact curve $S \subset \mathbb{T}$, which is given in our coordinates by
an equation
\begin{equation} \label{speccv}
P(\eta,\zeta) = \eta^{k}+
\alpha_{1}(\zeta)\eta^{k-1}+\ldots+\alpha_{k}(\zeta)=0
\end{equation}
where each $\alpha_{j}$ is a complex polynomial of degree not
exceeding $2j$. 
Notice that the real
structure $\tau$ induces antiholomorphic morphisms $L^{\pm}\xrightarrow
{\simeq} L^{\mp}$ 
and thus restricts to a real structure on $S$. This
implies that the polynomials $\alpha_{j}$ in equation (\ref{speccv})
must satisfy the reality conditions
\begin{equation} \label{real}
\alpha_{j}(\zeta)=(-1)^{j}\zeta^{2j}
\overline{\alpha_{j}(-\frac{1}{\bar{\zeta}})}.
\end{equation}
It can be shown that the three independent real coefficients of
$\alpha_{1}(\zeta)$ 
may be interpreted as giving the center $(x_{1},x_{2},x_{3})$ of the
monopole in $\mathbb{R}^{3}$,
\[ 
\alpha_{1}(\zeta)=k(x_{-}\zeta^{2}+2x_{3}\zeta -x_{+}) 
\]
where $x_{\pm}:=x_{1}\pm ix_{2}$, and are thus trivial moduli in the 
solution, related to the translational symmetry of (\ref{bgmln}). 
In the following, we shall only consider centred monopoles; these
are defined as having the origin as center and thus have 
$\alpha_{1}(\zeta)=0$.

In \cite{H2}, Hitchin proved that, conversely, any compact real curve
$S$ of the linear system $|{\cal O}(2k)|$ on $\mathbb{T}$ 
for which $L^{2}|_{S}$ is trivial determines
a charge $k$ monopole, which will be smooth if the additional condition
\begin{equation} \label{smooth}
H^{0}\left(S,L^{s}(k-2)\right)=0
\end{equation}
holds for $0<s<2$.
$S$ is called the spectral curve of the monopole and completely determines
the gauge equivalence class of the field configuration.
It encodes all the information about the monopole; in particular, its
genus $g$ is related to the magnetic charge $k$ by
\begin{equation} \label{genus}
g=(k-1)^2
\end{equation}
and every symmetry of $S$ is also a symmetry of the corresponding 
solution to~(\ref{bgmln}).

\section{A new version of the Ercolani--Sinha conditions} \label{ErcSin}

In \cite{ES}, Ercolani and Sinha rephrase the condition of triviality of the
line bundle $L^{2}|_{S}$ in terms of $g$ equations involving
periods of 1-forms on the spectral curve $S$. 
Starting with these equations, which they call the ``quantisation
conditions'', they propose an algorithm for constructing monopoles in
the case where the underlying spectral curve is nonsingular. 
We now review their argument.

Recall that when $S$ is nonsingular the group
$H^{0}(S,\Omega_{S}^{1})$ of global holomorphic 
1-forms on $S$ is a finite-dimensional $\mathbb{C}$-vector space, 
whose dimension is the genus $g$ of $S$.
Locally, these forms $\Omega$ can be described, using the adjunction 
formula, as 
Poincar\'e residues of meromorphic 1-forms on $\mathbb{T}$ 
with at most simple poles along $S$. Imposing global regularity, it is
easy to show that they can be written in our coordinates as
\begin{equation} \label{adjunction}
\Omega=\frac{\left( \beta_{0}\eta^{k-2}+\beta_{1}(\zeta)\eta^{k-3}+\ldots
+\beta_{k-2}(\zeta) \right) d\zeta}{\partial P / \partial \eta}
\end{equation}
(on $U_{0}\cap S$ and away from the branch points of $\pi|_{S}$), where 
each $\beta_{j}$ is a polynomial of degree at most $2j$ with arbitrary
coefficients.
It is clear from this formula that equation (\ref{genus}) indeed holds.

From equation (\ref{speccv}), it is clear that the spectral curve $S$
can be described as a $k$-sheeted branched cover of $\mathbb{CP}^{1}$, 
with projection $\pi|_{S}:S \rightarrow \mathbb{CP}^{1}$. 
The reality symmetry implies that the number of branch points is even
and that they occur in antipodal pairs. To define the sheets of the
cover, which we will label by integers $1,\ldots, k$,
we have to introduce appropriate branch cuts.
We may start by choosing a great circle on the
sphere passing through no branch points, and joining the branch points
in one of the corresponding hemispheres by non-intersecting
cuts; then we apply the antipodal map to these to get
further cuts joining the branch points on the other side of the
great circle we have chosen. To ensure
that each sheet is simply connected, we have to make one last cut,
connecting the cuts introduced on the two hemispheres. For the
spectral curves we shall consider below, one can argue that this last
cut has trivial monodromy and is thus unnecessary; in this
situation, the reality structure maps cuts to cuts and can therefore be
described in terms of the antipodal map together with an order two 
permutation of the sheets.

We will be interested in the local behaviour of 
certain meromorphic forms at the points of the fibre above $0$,
which we shall denote by $0_{j}$, $j = 1, \ldots, k$, and assume to be
distinct; this is no loss of generality since there is the freedom of
rotating the monopole.
Consider the meromorphic function on $S$ defined by 
${\eta}/{\zeta}$ on $U_{0}\cap S$; it is easy to see that it has
simple poles at the $2k$ points of $(\pi|_{S})^{-1}(\{0,\infty\})$
and is holomorphic elsewhere. 
In a neighbourhood of $0_{j}$,
\begin{equation} \label{defrho}
d\left( \frac{\eta}{\zeta}\right)=
\left(-\frac{\eta_{j}(0)}{\zeta^{2}}+O(1)\right)\,d\zeta
\;\;\;\;\;\;\;{\rm as}\;\zeta\rightarrow 0,
\end{equation}
where $\eta_{j}(\zeta)$ denotes the local solution of (\ref{speccv})
on the $j$th sheet. 
Given a global holomorphic 1-form $\Omega$, we introduce the 
notation $g_{j}$ for the coefficient of $\Omega$ at the point
$0_{j}$ in terms of the local coordinate $\zeta$, i.e.
\begin{equation} \label{defg}
\Omega|_{0_{j}}=:g_{j}\;d\zeta |_{\zeta=0}.
\end{equation}

The triviality of the line bundle $L^{2}|_{S}$ is
equivalent to the existence of a nowhere vanishing holomorphic section
$f$; with respect to the trivialisation of
$L^{2}|_{S}$ over the open sets $U_{0}\cap S$ and $U_{1}\cap S$, 
$f$ is given
by two nowhere vanishing holomorphic functions $f_{0}$ and $f_{1}$ on 
$U_{0}\cap S$, $U_{1}\cap S$ respectively, satisfying 
\[
f_{0}(\eta,\zeta)=e^{-2\frac{\eta}{\zeta}} f_{1}(\eta,\zeta)
\]
for $(\eta, \zeta) \in U_{0} \cap U_{1}\cap S$. This implies that the
meromorphic 1-forms $d\,{\rm log}f_{0}$ (:=$df_{0}/f_{0}$) 
and $d\,{\rm log}f_{1}$
are related by
\begin{equation} \label{dlog}
d\,{\rm log}f_{0}=-2d \left(\frac{\eta}{\zeta}\right) +d\,{\rm log} f_{1}
\end{equation}
on $U_{0} \cap U_{1}\cap S$. Notice that
 \begin{equation} \label{2piiZ}
 \oint _{\lambda} d\,{\rm log} f_{\alpha} \in 2 \pi i {\mathbb Z} \;\;\;\;\;\;
\alpha=0,1
\end{equation}
for any homology 1-cycle $\lambda \in H_{1}(U_{\alpha}\cap S,\mathbb{Z})$;
moreover, these integrals are nonzero in general, since the
1-forms $d\,{\rm log}f_{\alpha}$ do not have to be exact.
From equations (\ref{defrho}) and (\ref{dlog}), 
we conclude that $d\,{\rm log}f_{1}$ must have the local
behaviour near $0_{j}$
\[
d\,{\rm log}f_{1}=\left( -\frac{2 \eta_{j}(0)}{\zeta^{2}}+O(1)\right)\,d\zeta
\;\;\;\;\;\;\;{\rm as}\;\zeta\rightarrow 0
\]
in order for $f_{0}$ not to have an essential singularity at
$0_{j} \in U_{0}\cap S$.

It should be noted that the section $f$ is uniquely determined up to a
multiplicative constant, since the quotient of $f$ by any other
nowhere vanishing
section of $L^{2}|_{S}$ yields a global holomorphic function on the
compact Riemann surface $S$. Notice also that the modulus of
this constant can be
fixed by imposing the symmetry
\[
f_{1}(\eta,\zeta)=\frac{1}{\;\overline{f_{0} \circ \tau(\eta,\zeta)}\;}
\]
since the right-hand side has the regularity and nowhere vanishing 
properties of $f_{1}$, and $\frac{\eta}{\zeta}$ changes sign under
pull-back by $\tau$.

Let
$\{a_{1},\ldots,a_{g},b_{1},\ldots,b_{g}\}$
be a canonical basis of
$H_{1}(S,{\mathbb Z})\cong {\mathbb Z}^{\oplus 2g}$, i.e. 
satisfying the orthonormality conditions 
\begin{equation} \label{h1basis}
\sharp(a_{i},b_{j})=\delta_{ij} ,\;\;\;\;\;\;\; 
\sharp(a_{i},a_{j})=0= \sharp(b_{i},b_{j})
\end{equation}
for the intersection pairing.
Following Ercolani and Sinha, 
we apply the reciprocity law for differentials of the first and
second kinds (cf~\cite{GH}, p.~241) to an arbitrary 
holomorphic 1-form $\Omega$ and 
$d\,{\rm log}f_{1}$ to get 
\begin{equation} \label{quant1}
\sum_{i=1}^{k}(-2 \eta_{i}(0)) g_{i} 
= \frac{1} {2 \pi i} \sum_{j=1}^{g} \left|
\begin{array}{cc}
\oint_{a_{j}}\Omega & \oint_{a_{j}}d\,{\rm log}f_{1}\\
\oint_{b_{j}}\Omega & \oint_{b_{j}}d\,{\rm log}f_{1}
\end{array} \right|.      
\end{equation}
Let $m_{j}$ and $n_{j}$ be the integers
\begin{equation}\label{defmn}
m_{j}:=-\frac{1}{2 \pi i} \oint_{a_{j}}d\,{\rm log} f_{1}
\;\;\; {\rm and} \;\;\;
n_{j}:=\frac{1}{2 \pi i} \oint_{b_{j}}d\,{\rm log} f_{1},
\end{equation}
consistently with (\ref{2piiZ}), and let us define the 1-cycle
\begin{equation}\label{ESc}
c:=\sum_{j=1}^{g}(n_{j}a_{j}+m_{j}b_{j}).
\end{equation}
Then equation (\ref{quant1}) can be rewritten as
\begin{equation} \label{quant2}
- 2 \sum_{i=1}^{k} \eta_{i}(0) g_{i} =  
\oint_{c}\Omega.
\end{equation}

The existence of $c \in H_{1}(S,\mathbb{Z})$ satisfying
(\ref{quant2}) is equivalent to the line bundle $L^{2}|_{S}$ being
trivial. Unfortunately, the condition~(\ref{smooth}) which would
ensure smoothness cannot be implemented directly in the
Ercolani--Sinha approach if $k>2$, but
we can include a weaker statement in the analysis as follows.
Since for $k \ge 2$ there is an 
inclusion $H^{0}(S,L^{s})\hookrightarrow H^{0}(S,L^{s}(k-2))$
given by tensoring with a section of ${\cal O}(k-2)|_{S}$,
the condition 
\begin{equation} \label{weaker}
H^{0}(S,L^{s})=0 
\end{equation}
is necessary for (\ref{smooth}) to
hold. Now we can repeat the argument above to investigate the existence
of global sections of $L^{s}|_{S}$, arriving at the same equation
(\ref{quant2}) with $2$ replaced by~$s$, and we can conclude that
there will be no nontrivial global sections of $L^{s}|_{S}$ for $0<s<2$ 
if and only if $c$ is 
primitive in $H_{1}(S,\mathbb{Z})$.

We can still simplify the left-hand side of (\ref{quant2}).
Consider a global holomorphic 1-form $\Omega$ on $S$, as given by 
(\ref{adjunction}). After defining the branch cuts, we can write
\begin{equation} \label{factp}
P(\eta, \zeta)=\prod_{j=1}^{k} \left( \eta - \eta_{j}(\zeta) \right),
\end{equation}
and so
\[
\frac{\partial P}{\partial \eta} (\eta, \zeta) = \sum_{i=1}^{k}
\prod_{j\ne i}^{k} \left(\eta-\eta_{j}(\zeta) \right).
\]
On sheet $i$, $\eta=\eta_{i}(\zeta)$ and all the terms in the sum 
above vanish except one,
\begin{equation}\label{dpdej}
\left.\frac{\partial P}{\partial \eta}(\eta,\zeta)\right|_{\text{sheet }i} =
\prod_{j\ne i}^{k}\left(\eta_{i}(\zeta)-\eta_{j}(\zeta)\right).
\end{equation}
We can use this to write the coefficient $g_{i}$ in (\ref{defg}) for
$\Omega$ as
\[
g_{i}=\frac{\beta_{0}\eta_{i}^{k-2}(0)
+\beta_{1}(0)\eta_{i}^{k-3}(0)+ \ldots+ \beta_{k-2}(0)}
{ \prod_{j \neq i}^{k} \left( \eta_{i}(0)-\eta_{j}(0)\right)},
\]
so the left-hand side of (\ref{quant2}) takes the form
\[
-2\sum_{i=1}^{k}\eta_{i}(0)g_{i}=-2\sum_{i=1}^{k}
\frac{\beta_{0}\eta_{i}^{k-1}(0)+\beta_{1}(0)\eta_{i}^{k-2}(0)+
\ldots+\beta_{k-2}(0)\eta_{i}(0)}
{ \prod_{j\neq i}^{k}\left( \eta_{i}(0)-\eta_{j}(0) \right)}.
\]
This appears to be a very complicated expression, but we can
simplify it considerably if we make use of the identity
\begin{equation}\label{niceid}
\sum_{i=1}^{k} x_{i}^{n} \prod_{j\ne i}^{k}\frac{1}{x_{i}-x_{j}}=
\begin{cases}
0& ,0 \le n \le k-2 \\
1& ,n=k-1
\end{cases}.
\end{equation}
Taking $x_{i}=\eta_{i}(0)$, we obtain
\begin{equation}\label{simple}
-2\sum_{i=1}^{k}\eta_{i}(0)g_{i}=-2 \beta_{0}
\end{equation}
and substitution in (\ref{quant2}) yields
\begin{equation}\label{ourES}
\oint_{c}\Omega = -2\beta_{0}.
\end{equation}
So our version of the Ercolani--Sinha conditions amounts to the 
existence of a primitive 1-cycle $c$ such that equation (\ref{ourES}) 
is satisfied for every global holomorphic 1-form $\Omega$, 
where $\beta_{0}$ is the coefficient in (\ref{adjunction}) for $\Omega$.

To prove (\ref{niceid}), we first note that the cases $0 \le n \le
k-2$ follow from the $n=k-1$ case: A translation $x_{i} \mapsto 
x_{i}-y$ of all the $x_{i}$'s leaves the denominators in the sum
invariant,
\[
\sum_{i=1}^{k} (x_{i}-y)^{k-1} \prod_{j\ne i}^{k}\frac{1}{x_{i}-x_{j}}=1,
\]
so by expanding the binomials and collecting equal powers of $y$ we get
the statement for all $0 \le n \le k-2$. The proof of the $n=k-1$ case
by induction on $k$ is rather lengthy and we prefer to argue as follows. 
It is readily seen that the whole sum is symmetric under the action of 
the symmetric group $\mathfrak{S}_{k}$ permuting the
$x_{i}$'s. Reducing 
to a common fraction yields as denominator
\[
\Delta(x_{1},\ldots,x_{k})=\prod_{i<j}^{k}(x_{i}-x_{j})
\]
and this polynomial is completely antisymmetric under
$\mathfrak{S}_{k}$; in fact, the space of antisymmetric
polynomials in $k$ variables is generated by $\Delta$ over the ring of
symmetric polynomials. The numerator is then necessarily 
antisymmetric and a homogeneous polynomial of degree
$\frac{1}{2}k(k-1)$, which is
also the degree of $\Delta$, so it has to be equal to $\Delta$ times 
a constant. Taking the asymptotic limit $x_{1} \rightarrow \infty$
in the original sum, we conclude that this constant has to be 1.

It is convenient, when we come to investigate particular examples, to
introduce bases for both the global holomorphic 1-forms and the
homology 1-cycles on $S$.
An obvious basis $\{\Omega^{(\ell)},1\le
\ell \le g\}$ for $H^{0}(S,\Omega_{S}^{1})$ corresponds 
to taking monomials $\eta^{r} \zeta^{s}$ for
the allowed powers $r$ and $s$ (in lexicographical order of decreasing
$r$ and increasing $s$) as numerators of (\ref{adjunction}),
\begin{equation}\label{obvbasis}
\Omega^{(1)}=\frac{\eta^{k-2}d\zeta}{\partial P / \partial \eta},\;\;
\Omega^{(2)}=\frac{\eta^{k-3}d\zeta}{\partial P / \partial \eta},\;\; 
\Omega^{(3)}=\frac{\eta^{k-3}\zeta d\zeta}{\partial P / \partial \eta},\;\; 
\ldots,\;\;
\Omega^{(g)}=\frac{\zeta^{2k-4}d\zeta}{\partial P / \partial \eta}.
\end{equation}
The condition (\ref{ourES}) for a general $\Omega$ is then equivalent to
the $g$ conditions
\begin{equation}\label{justleq1}
\oint_{c}\Omega^{(\ell)}=-2\delta_{1 \ell}.
\end{equation}   
Let us also fix a canonical basis (\ref{h1basis}) for $H_{1}(S,\mathbb{Z})$.
The ($g \times 2g$) period matrix for $S$ corresponding to the two
choices of bases is then defined as usual by $\mathbf{P}=[\mathbf{A}|
\mathbf{B}]$, where $\mathbf{A}$ and $\mathbf{B}$ are square matrices 
with entries
\[
A_{\ell j}:=\oint_{a_{j}}\Omega^{(\ell)} \;\;\;\;{\rm and} \;\;\;\;
B_{\ell j}:=\oint_{b_{j}}\Omega^{(\ell)}.
\]
Recalling (\ref{ESc}), equation (\ref{justleq1}) can now be written as
\begin{equation} \label{explES}
\sum_{j=1}^{g} (A_{\ell j}n_{j}+B_{\ell j}m_{j})=-2\delta_{1\ell}.
\end{equation}

Although the number of integers to be determined in (\ref{explES}) is 
$2g$, they still have to satisfy constraints coming from the reality 
structure of $S$. 
We prove below that these imply that $c$ is antisymmetric under the
action of $\tau$ on the first homology group,
\begin{equation}\label{casym}
\tau_{*}c=-c.
\end{equation}
This imposes $g$ linear constraints on the $2g$ components of $c$. 
In fact, since $\tau$ is anti-holomorphic,
\[
\sharp(a,b)=-\sharp(\tau_{*}a,\tau_{*}b)
\]
for any $a,b \in H_{1}(S,\mathbb{Z})$, and this shows that the matrix
$\boldsymbol{\tau}$ representing $\tau_{*}$ in the canonical basis
(\ref{h1basis}) of $H_{1}(S,\mathbb{Z})$ satisfies
\begin{equation}\label{almostsymp}
\boldsymbol{\tau}^{t}=\mathbf{J}(-\boldsymbol{\tau}^{-1})\mathbf{J}^{-1}
\end{equation}
where $\mathbf{J}$ is the matrix representing the intersection
pairing in this basis,
\begin{equation}\label{jmatrix}
\mathbf{J}=\begin{bmatrix}
 & \mathbf{1}_{g} \\
-\mathbf{1}_{g} & 
\end{bmatrix}.
\end{equation}
Since $\boldsymbol{\tau}^{2}=\mathbf{1}_{2g}$, $\boldsymbol{\tau}$ is
diagonalisable and has eigenvalues $\pm 1$; then (\ref{almostsymp}) implies
that these have to occur with equal multiplicities. Hence the
antisymmetric 1-cycles lie in a $\mathbb{Z}^{\oplus g}$ subgroup of
$H_{1}(S,\mathbb{Z})$.

To prove (\ref{casym}), we consider the basis (\ref{obvbasis}). 
Since $\tau$ is antiholomorphic, it pulls back holomorphic 1-forms on 
$S$ to antiholomorphic 1-forms and vice-versa; the forms above are
mapped as
\begin{equation}\label{pbtau}
\tau^{*}\left(\frac{\eta^{r}\zeta^{s}d\zeta}{\D {\partial
P}/{\partial\eta}(\eta,\zeta)}\right)=
(-1)^{k+r+s+1}\overline{\left( \frac{\eta^{r} 
\zeta^{2(k-r-2)-s}d\zeta}
{\D {\partial P}/{\partial \eta}(\eta,\zeta)}\right)}
\end{equation}
for $0\le r \le k-2$ and $0\le s \le 2(k-r-2)$.
Using (\ref{pbtau}) and (\ref{justleq1}), we obtain
\[
\oint_{\tau_{*}c+c}\Omega^{(1)}=
-\overline{\oint_{c}\Omega^{(1)}}+\oint_{c}\Omega^{(1)}=2-2=0
\]
and for $\ell \ne 1$
\[
\oint_{\tau_{*}c+c}\Omega^{(\ell)}=
\pm \overline{\oint_{c}\Omega^{(\ell')}}+\oint_{c}\Omega^{(\ell)}=
\pm 0+0=0
\]
for some $\ell' \ne 1$. We conclude that the integral of any global
holomorphic 1-form around $\tau_{*}c+c$ vanishes, and this implies
(\ref{casym}).

To illustrate how we can use the conditions (\ref{explES}) to
determine spectral curves of monopoles, we take as example the
well-known charge 2 monopole (\cite{Hur}, \cite{AH}), which is also 
considered in \cite{ES}. The general spectral curve for a centred
monopole of charge 2, after imposing the reality conditions
(\ref{real}), has the form
\[
\eta^2 +
(\gamma_{0}\zeta^4+\gamma_{1}\zeta^3+\gamma_{2}\zeta^2-\overline{\gamma}_{1}
\zeta+\overline{\gamma}_{0})=0
\]
where $\gamma_{2}$ is real. The four roots of the polynomial in brackets
occur in antipodal pairs; we can use the $SO(3)$ action to take one
pair to $\pm 1$ and the other one to
$\pm e^{\pm 2\theta i}$, where $0 \le \theta \le \frac{\pi}{4}$. A further
rotation by $\zeta \mapsto e^{i\theta} \zeta$ then takes the spectral
curve to
\begin{equation} \label{chrg2}
\eta^2 + \left( \frac{\kappa}{2}\right)^{2} \left( \zeta^4-2 \cos
(2\theta)
\zeta^2 + 1\right)=0
\end{equation}
where $\kappa$ is a real number to be determined in terms of $\theta$.

Equation (\ref{chrg2}) defines a double cover of $\mathbb{CP}^{1}$
with branch points at the four roots of the polynomial in brackets,
\[
w_{1}=e^{i\theta},\;\; 
w_{2}=-e^{-i\theta},\;\;
z_{1}=-e^{i\theta},\;\;
z_{2}=e^{-i\theta}.
\]
We will be interested in the generic case where $S$ is nonsingular; this
happens if and only if all the points above are distinct. $S$ is an
elliptic curve and can be constructed by gluing together two copies
of the Riemann sphere along two branch cuts, that we choose to be on
the equator
$\{\zeta:|\zeta|=1\}$. 
\begin{figure}[ht]
\begin{center}
\epsfig{file=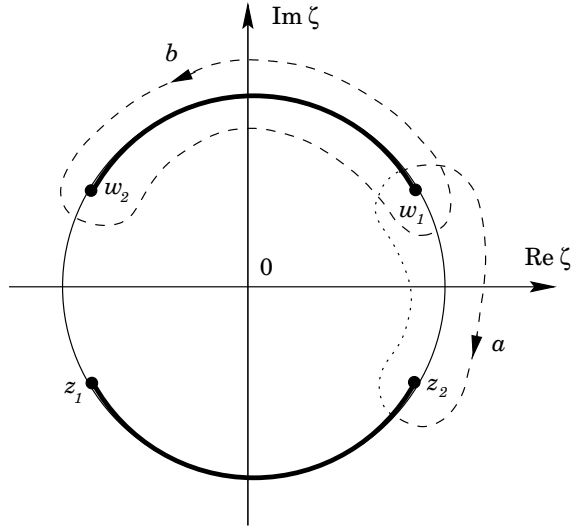}
\caption{{\small \it Branch cuts and 1-homology basis for the spectral
curve of the charge $2$ monopole}} 
\label{figchrg2}
\end{center}
\end{figure}
We label the two sheets of $S$ by $j=1,2$, which correspond to the two
possible choices of sign for $\eta$ when solving
(\ref{chrg2}); sheet $j$ is defined by the function $\eta_{j}$
obtained by analytic
continuation, avoiding the cuts above, of
\[
\zeta \mapsto(-1)^{j-1} \frac{i \kappa}{2} \sqrt{\zeta^4-2\cos (2\theta)
\zeta ^2+1}
\]
regarded as a germ at $0\in {\mathbb C}$. Here, and elsewhere, we
consider the principal branch of the root, viz
$-\frac{\pi}{q} < \arg {z}^{1/q} \le \frac{\pi}{q}, \; \forall \,z \in
{\mathbb C}^{\,*}$.

We choose a canonical basis $\{a,b\}$ of $H_{1}(S,\mathbb{Z})$ 
as in Figure~\ref{figchrg2}, where we draw the paths as dashed
or dotted lines if they lie on sheets $1$ or $2$,
and write $c=na+mb$.
In this case $H^{0}(S,\Omega_{S}^{1})$ is 1-dimensional and a generator is
\[
\Omega=\frac{d\zeta}{2 \eta}.
\]
The periods can be expressed in terms of Legendre's complete 
elliptic integral of the first kind,
\[
A=\oint_{a}\Omega= \frac{2}{\kappa \sin \theta}\int_{0}^{\theta}
\frac{du}{\sqrt{1-\csc^{2}\theta \sin^{2}u}}=
\frac{2}{\kappa}K(\sin\theta)
\]
\[
B=\oint_{b}\Omega=\frac{2i}{\kappa \cos \theta}\int_{0}^{\frac{\pi}{2}-\theta}
\frac{du}{\sqrt{1-\sec^{2}\theta \sin^{2}u}}=
\frac{2i}{\kappa}K(\cos\theta).
\]
So equation (\ref{explES}) reads
\[
\frac{2}{\kappa}K(\sin \theta)n+\frac{2i}{\kappa}K(\cos \theta)m
=-2.
\]
Therefore $m=0$, and $n$ must then be a generator of $\mathbb Z$, which we can
take to be $-1$, obtaining
\[
\kappa=K(\sin\theta).
\]
This can be checked to agree with the result of Hurtubise
\cite{Hur}. Note that in this case equations (\ref{smooth}) and
(\ref{weaker}) are equivalent, so the method recovers all nonsingular 
spectral curves of (centred and suitably oriented) monopoles of charge
$2$.

In this example, the special 1-cycle~$c$ in equation~(\ref{ourES}) is
thus~$-a$. It is readily checked that it is antisymmetric under $\tau$.
We point out that, although here
\[
\oint_{a}d\log f_{1}=0,
\]
the $a$-periods of $d\log f_{1}$ do not vanish for general monopoles, and 
this cannot be avoided by just rescaling  $f$ as claimed in~\cite{ES}. 
This will be illustrated in section~\ref{4hedralspec}, where we
consider a monopole with a spectral curve of higher genus.

\section{The Corrigan--Goddard conditions} \label{CorGod}

In \cite{CG}, Corrigan and Goddard used the so-called ${\cal A}_{k}$
Ansatz of Atiyah--Ward for instantons to construct a charge $k$
solution to the Bogomol'ny\u\i\ equations (\ref{bgmln}) with 
$\dim {\cal N}_{k}=4k-1$ free parameters. This construction was also 
obtained independently by Forg\'acs et al. \cite{FHP}, and has been 
applied \cite{ORS} to study monopoles in situations where the
equations involved are simplified.
Unlike the method we presented in section \ref{ErcSin}, the Corrigan--Goddard
approach does not assume smoothness of the underlying spectral curves;
indeed, it can be used to obtain for example the axially symmetric monopole of
arbitrary charge $k$, whose spectral curve is reducible to $k$ spherical
components.

In the notation we have introduced, the construction goes as follows. 
Start with a polynomial $P(\eta,\zeta)$ as in (\ref{speccv}), satisfying the
reality constraints (\ref{real}).
Orient the monopole so that there is an open annulus $A$ in
$\mathbb{CP}^{1}$ which contains the equator $E=\{\zeta:|\zeta|=1\}$ but
does not contain any of the branch points of $\pi|_{S}$.
Assume that $A$ lifts to $k$ disjoint annuli on the spectral curve;
then one can define the branch cuts so that sheet $j$ contains one
of the lifted annuli, which we denote by $A_{j}$.
On $\pi^{-1}(A)$, consider the function
\begin{equation} \label{bigtheta}
\Theta (\eta, \zeta):=2 \pi i \sum_{j=1}^{k}\frac{\nu_{j}}{2}
\prod_{\ell \ne j}^{k} \frac{\eta-\eta_{\ell}(\zeta)}
{\eta_{j}(\zeta)-\eta_{\ell}(\zeta)}
\end{equation}
where $\nu_{j}$ are some integers to be determined. This is
a Lagrange interpolation polynomial in $\eta$ of the $k$ conditions that
$\Theta$ should take the value $\pi i \nu_{j}$ on $A_{j}$.
For $\zeta \in A$, define the functions $\Theta_{r}$ from the
coefficients of $\eta^{r}$ in $\Theta$ as follows:
\begin{equation}\label{thetar}
\Theta(\eta,\zeta)=: 2\pi i \sum_{r=0}^{k-1} \Theta_{r}(\zeta) 
\left(\frac{\eta}{2\zeta}\right)^{r}.
\end{equation}
Corrigan and Goddard's analysis then leads to the conditions
\begin{equation}\label{CGtheta1}
\oint_{E}\Theta_{1}(\zeta) \frac{d\zeta}{\zeta} = 2
\end{equation}
and
\begin{equation}\label{CGthetar}
\oint_{E}\Theta_{r}(\zeta)\zeta^{s} \frac{d\zeta}{\zeta}=0,\;\;\;2\le r \le
k-1,\; |s| \le r-1.
\end{equation}
These are $(k-1)^{2}$ constraints on the $k^{2} + 2k$ coefficients of
$P(\eta,\zeta)$, just as one obtains using the Ercolani--Sinha
algorithm. When the
spectral curve is nonsingular, we would expect them to be equivalent
to (\ref{justleq1}). We now clarify how they relate to each other.

Denoting by $s_{i}$ the $i$th elementary symmetric polynomial
in a given number of variables, we can expand the numerator of 
(\ref{bigtheta}) to obtain
\[
\Theta(\eta,\zeta)=2\pi i \sum_{r=0}^{k-1}\sum_{j=1}^{k}
(-1)^{k-r-1}\frac{\nu_{j}}{2}\frac{s_{k-r-1}(\eta_{1}(\zeta),
\ldots,\widehat{\eta_{j}(\zeta)},\ldots,\eta_{k}(\zeta))\eta^{r}}
{\prod_{\ell \ne j}^{k}\left(\eta_{j}(\zeta)-\eta_{\ell}(\zeta) \right)}.
\]
The elementary symmetric polynomials satisfy the recurrence relation
\[
s_{i}(x_{1},\ldots,\widehat{x_{j}},\ldots,x_{k})=s_{i}(x_{1},\ldots,x_{k})-
x_{j}s_{i-1}(x_{1},\ldots,\widehat{x_{j}},\ldots,x_{k})
\]
for $0\le i \le k$ (taking $s_{0}:=1$), and iterating this one finds
\[
s_{i}(x_{1},\ldots,\widehat{x_{j}},\ldots,x_{k})=
\sum_{h=0}^{i}(-1)^{h}x_{j}^{h}s_{i-h}(x_{1},\ldots,x_{k}).
\]
Clearly, $(-1)^{j}s_{j}(\eta_{1}(\zeta),\ldots,\eta_{k}(\zeta))$ are
just the polynomials $\alpha_{j}(\zeta)$ in (\ref{speccv}) for each $0 \le i
\le k$ (with $\alpha_{0}:=1$). Therefore, we can read off
the functions $\Theta_{r}$ in (\ref{thetar}) as
\[
\Theta_{r}(\zeta)=\sum_{j=1}^{k}\sum_{h=0}^{k-r-1}\frac{\nu_{j}}{2}
\frac{\eta_{j}^{h}(\zeta)\alpha_{k-r-h-1}(\zeta)}
{\prod_{\ell \ne j}^{k}\left( \eta_{j}(\zeta)
-\eta_{\ell}(\zeta)\right)}(2\zeta)^{r}.
\]
So far, we have shown that, for the $j$th term in the sum, the
numerator depends only on $\eta_{j}(\zeta)$ and $\zeta$.
Using (\ref{dpdej}), we can eliminate altogether the dependence
on the functions $\eta_{\ell}$ with $\ell\ne j$, and this allows 
us to write for $1\le r \le k-1$
\[
\oint_{E}\Theta_{r}(\zeta)\frac{d \zeta}{\zeta}=
{ \oint}_{\sum_{j=1}^{k}\nu_{j} E_{j}}
\Bigl.
\frac{\D \sum_{h=0}^{k-r-1}\eta^{h}
\alpha_{k-r-h-1}(\zeta)}{\D \frac{\partial P}{\partial
\eta}(\eta,\zeta)}
\,(2\zeta)^{r-1}\,d\zeta
\Bigr. ,
\]
where $E_{j}:= (\pi|_{S})^{-1}(E)\cap A_{j}$ is the lift of $E$ to
sheet $j$. The integrand no longer depends on the sheet label.
It becomes clear now how to write the left-hand side of the 
Corrigan--Goddard conditions
as integrals over 1-cycles on $S$. If we define the holomorphic 1-form 
$-\Xi_{r}$ on $\cup_{j=1}^{k} A_{j}$ to be the integrand in the above
expression, then the conditions (\ref{CGtheta1}) and (\ref{CGthetar}) can
be written respectively as
\begin{equation}\label{newCG1}
\oint_{\sum_{j=1}^{k}\nu_{j}E_{j}}\Xi_{1}=-2
\end{equation}
and
\begin{equation}\label{newCGr}
\oint_{\sum_{j=1}^{k}\nu_{j} E_{j}} \zeta^{s}\:\Xi_{r}=0
\end{equation}
for $2\le r \le k-1$ and $|s|\le r-1$.

Equations (\ref{newCG1}) and (\ref{newCGr}) are very similar to the
version (\ref{justleq1}) of the Ercolani--Sinha conditions. 
In fact, they turn out to be precisely 
equivalent to (\ref{justleq1}), provided
we assume $c$ to be of the form
\begin{equation}\label{CGc}
c=\sum_{j=1}^{k} \nu_{j} E_{j}
\end{equation}
rather than a general 1-cycle as in (\ref{ESc}).
To see this, we first remark that all the integrands in (\ref{newCG1})
and (\ref{newCGr}) are of the form (\ref{adjunction}), and hence global
holomorphic 1-forms on $S$. For each $1\le r \le k-1$, the highest
power of $\eta$ in the numerator of $\Xi_{r}$ never exceeds $k-r-1$,
and the coefficient of $\eta^{k-r-1}$ can be seen to be equal to 
$-(2\zeta)^{r-1}$. So multiplication of $\Xi_{r}$ by $\zeta^{s}$ 
with $-r+1\le s\le r-1$ as in (\ref{newCGr}) 
gives monomials in $\zeta$ of all degrees between $0$ and $2(r-1)$ as 
coefficients for $\eta^{k-r-1}$. 
We conclude that all the
homogeneous equations ($\ell \ne 1$) in (\ref{justleq1}) 
can be obtained from
(\ref{newCGr}) if we consider first the $2k-3$ equations
corresponding to $r=k-1$ and continue decreasing $r$ down to $2$, 
using at each stage the vanishing of the integrals for greater $r$
from the previous steps.
The $\ell=1$ equation also agrees with (\ref{newCG1}), since we
can use (\ref{newCGr}) and the coefficient of $\eta^{k-2}$ in the
numerator of $\Xi_{1}$ is $-1$. Conversely, the Ercolani--Sinha
conditions in the form (\ref{justleq1}) also imply the 
Corrigan--Goddard conditions (\ref{newCG1}) and (\ref{newCGr}) if
(\ref{CGc}) holds.

The question to put now is of course: Is the Ansatz (\ref{CGc}) for
the special cycle $c$ in equation (\ref{ourES}) valid in general? 
In the next section, we show that this is not the case, by explicit
computation of $c$ for the tetrahedral 3-monopole. 

\section{The tetrahedral 3-monopole revisited} \label{4hedral}

\subsection{Spectral curve} \label{4hedralspec}

Now we apply the method of section \ref{ErcSin} to investigate the 
spectral curve of the tetrahedrally symmetric monopole of charge~$3$. 
This was first 
studied in~\cite{HMM}, where the existence of the monopole was proved by
imposing tetrahedral symmetry to simplify Nahm's equations and solve them
in terms of elliptic functions. A numerical treatment of the ADHMN
construction was developed and applied to this monopole in~\cite{HS}, 
which allowed the fields to be computed and, using these, level
surfaces for the energy density were plotted.

\begin{figure}[ht]
\begin{center}
\epsfig{file=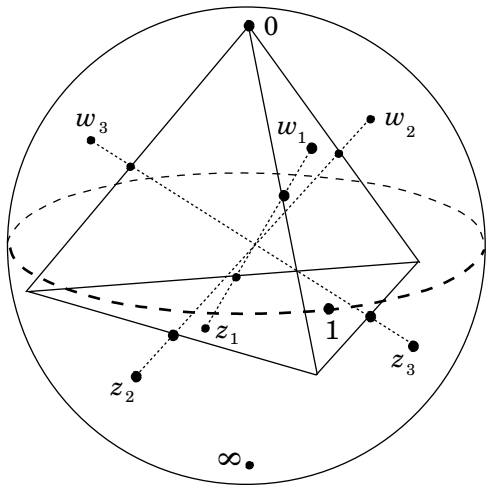,width=6.5cm}
\caption{{\small \it The inscribed tetrahedron underlying the symmetry
of the spectral curve}} 
\label{figvrtxup}
\end{center}
\end{figure}

As in \cite{HMM}, we start with the Ansatz
\begin{equation} \label{ch3ansatz}
\eta^{3}+\alpha(\zeta^{6}+5\sqrt{2}\zeta^{3}-1)=0
\end{equation}
for the spectral curve $S$, where $\alpha$ is a nonzero 
constant to be determined; the reality conditions imply $\alpha \in
\mathbb{R}$. 
The branch points occur at the zeroes of the polynomial in brackets,
\[
w_{1}=\frac{\sqrt{3}-1}{\sqrt{2}}, \;w_{2}=\omega w_{1},
\;w_{3}=\bar{\omega} w_{1}, 
\;z_{1}=-\frac{\sqrt{3}+1}{\sqrt{2}}, \;z_{2}=\omega z_{1},
\;z_{3}=\bar{\omega} z_{1},
\]
where $\omega:=e^{\frac{2\pi i}{3}}$. These are equidistant points on
the Riemann sphere, antipodal in pairs, which are related by radial
projection to the midpoints of the edges of a tetrahedron inscribed in
the sphere. In the configuration we have chosen, the tetrahedron has a
vertex at $0$ and is oriented such that the radial projection of one
of the three edges containing $0$ passes through $1$, as shown in
Figure~\ref{figvrtxup}.

\begin{figure}[ht]
\begin{center}
\epsfig{file=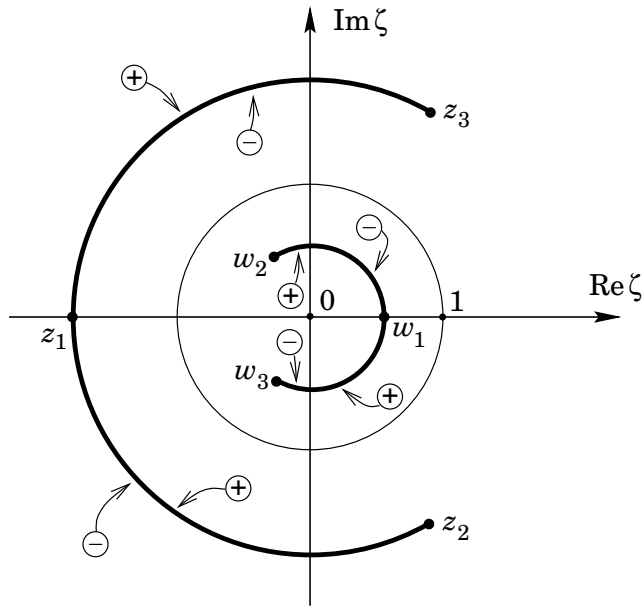,width=8.5cm}
\caption{{\small \it Branch cuts for the spectral curve of the tetrahedral 
3-monopole}} 
\label{figcuts}
\end{center}
\end{figure}

To define the branch cuts, we choose to connect the $w_{i}$'s and the 
$z_{i}$'s together along arcs of circles centred at the origin and
antipodal to each other as shown in Figure~\ref{figcuts}. No
more cuts are needed, since each branch point is of cube root type
and so any closed path on $\mathbb{CP}^{1}$ enclosing zero mod 3 branch
points lifts to a closed path on $S$.
Now we can label the three sheets as before: for $j=1,2,3$, we define
sheet $j$ to correspond to the analytic continuation $\eta_{j}$ of
\begin{equation} \label{etaj}
\zeta \mapsto - \omega^{j-1} \alpha^{1/3}\, \zeta \sqrt[3]{\zeta^{3} + 5
\sqrt{2}-\zeta^{-3}} 
\end{equation}
regarded as a germ at $1 \in {\mathbb C}$. In particular, notice that
on each sheet $\eta$ is indeed given by (\ref{etaj}) for all $\zeta$ in
the annulus $C:=\{\zeta:|w_{1}|< | \zeta|  <|z_{1}|\}$. With these 
conventions, it can be checked
that the rules for crossing the branch cuts are as given
in Figure~\ref{figcuts}, where the encircled $\pm$ signs mean that the
label $j$ is to be increased/decreased by $1$ mod $3$ when the corresponding
cut is crossed.

\begin{figure}[ht]
\begin{center}
\epsfig{file=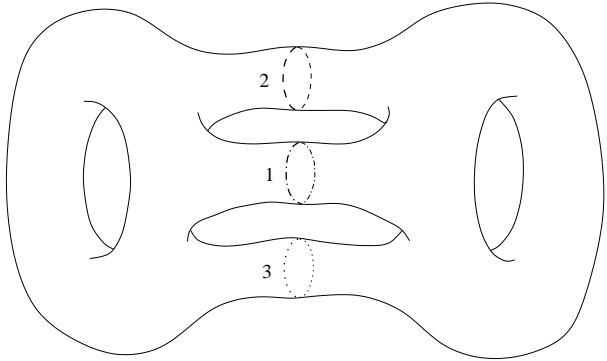,width=8cm}
\caption{{\small \it Spectral curve of the tetrahedral 3-monopole}}
\label{figspech3}
\end{center}
\end{figure}

It is not hard to see that one obtains a compact Riemann surface
of genus four when three copies of the Riemann sphere are identified 
along the branch cuts as
specified in Figure~\ref{figcuts}. In fact, by identifying three 
copies of the upper or lower hemispheres along the pair of cuts
as above, one obtains a torus with three discs removed; the circles of the
boundary correspond to the equators of the spheres we started
with. Gluing together the two surfaces obtained in this way along their
boundaries gives a compact curve of genus four. This is sketched 
in Figure~\ref{figspech3}; the three circles shown project under $\pi|_{S}$
to the equator $E$ of $\mathbb{CP}^{1}$, and they will
be
referred to as the equators on a given sheet. We shall adopt the convention
of drawing the paths as dash-dotted, dashed or dotted curves if they
lie on sheets $1$, $2$ or $3$, respectively.

\begin{figure}[ht]
\begin{center}
\epsfig{file=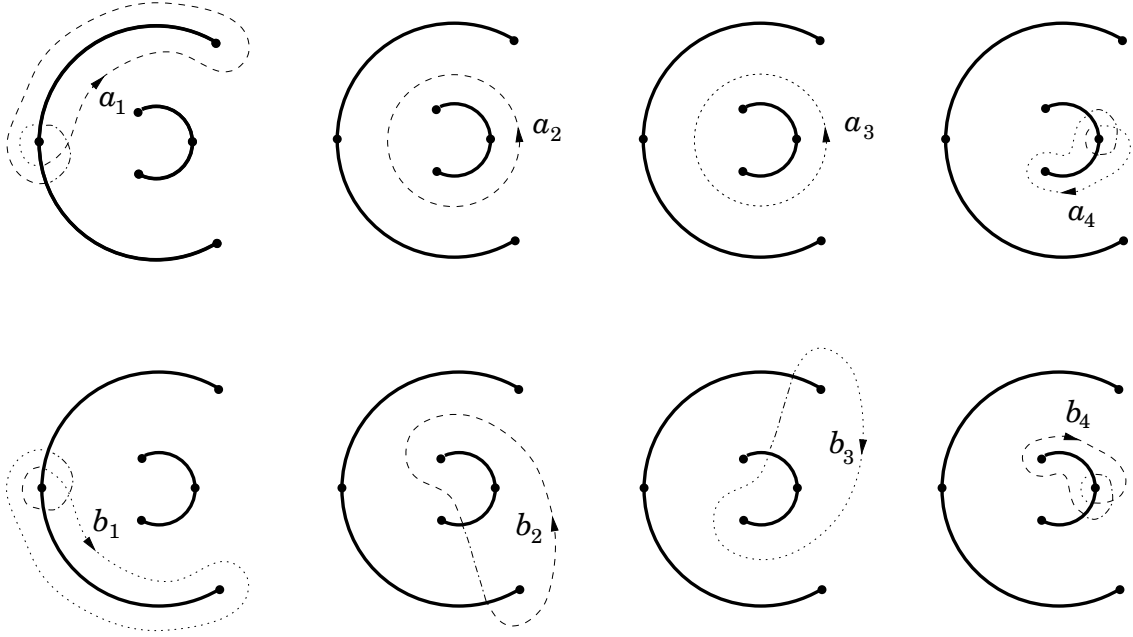,width=15cm}
\caption{{\small \it The basis for $H_{1}(S,\mathbb{Z})$}}
\label{fighbasis}
\end{center}
\end{figure}

Now we choose a canonical basis for
$H_{1}(S,\mathbb{Z})\cong\mathbb{Z}^{\oplus8}$ as in
Figure~\ref{fighbasis}. The first
two $1$-cycles $a_{1}$ and $b_{1}$ are drawn close to the cut 
connecting the $z_{i}$'s so as to have the desired
intersection number; for $a_{2}$ and $b_{2}$ we choose the equator 
on sheet 2 and a
distorted meridian intersecting it as required; all the
other intersections between these four $1$-cycles are zero. Then we
act with the reality map $\tau$ on these cycles to get the other
elements of the basis: 
\begin{equation} \label{nicebasis}
a_{3}:=\tau_{*}a_{2},\;\; a_{4}:=\tau_{*} a_{1},\;\;
b_{3}:=-\tau_{*} b_{2},\;\; b_{4}:=-\tau_{*}b_{1}. 
\end{equation}
Our choice of
branch cuts is such that $\tau$ sends cuts to cuts and hence maps a 
given sheet onto another sheet. 
It is easy to check that for $\zeta \in \mathbb{R}$, $\eta$
as given by (\ref{etaj}) for $j=1$ also takes real values
(cf equation (\ref{tau})). 
We then conclude that sheet $1$ is 
invariant under $\tau$, while the other two sheets are interchanged. 
It follows that the second half of our homology basis is
as drawn in Figure~\ref{fighbasis}, and all the remaining 
intersection numbers for the elements in the basis are as required 
by~(\ref{h1basis}). 

Our chosen basis for $H^{0}(S,\Omega_{S}^{1})$ is
\begin{equation}\label{omega14} 
\Omega^{(1)}=\frac{d\zeta}{3 \eta},\;\;
\Omega^{(2)}=\frac{d\zeta}{3 \eta^{2}},\;\;
\Omega^{(3)}=\frac{\zeta d\zeta}{3 \eta^{2}},\;\;
\Omega^{(4)}=\frac{\zeta^{2} d\zeta}{3 \eta^{2}}.
\end{equation}
According to (\ref{pbtau}) these forms are pulled back 
by the reality structure as
\begin{equation} \label{pbbytau}
\tau^{*}\Omega^{(1)}=-\overline{\Omega^{(1)}},\;\;\;\;
\tau^{*}\Omega^{(2)}=\overline{\Omega^{(4)}},\;\;\;\;
\tau^{*}\Omega^{(3)}=-\overline{\Omega^{(3)}},\;\;\;\;
\tau^{*}\Omega^{(4)}=\overline{\Omega^{(2)}}.
\end{equation}

We are now ready to compute the period matrix. The reality properties
(\ref{nicebasis}) and (\ref{pbbytau}) imply that
the periods around $a_{2}$, $a_{1}$,
$b_{2}$ and $b_{1}$ determine those around $a_{3}$, $a_{4}$,
$b_{3}$ and $b_{4}$, respectively. For example,
\begin{equation} \label{halfof32}
B_{23}=\oint_{-\tau_{*}b_{2}}\overline{\tau^{*}\Omega^{(4)}}=
-\overline{\oint_{b_{2}}\Omega^{(4)}}=-\overline{B_{42}}.
\end{equation}
This means that we only have to
calculate half of the $32$ entries of the period matrix.

First we consider the periods around the equator $a_{2}$. Notice that
$\zeta^{3}+5\sqrt{2}-\zeta^{-3}$ is invariant under the change of
variable $\zeta \mapsto \omega \zeta$. So
\[
A_{22}=-\frac{(1+\omega+\bar{\omega})}{3 \alpha^{1/3}}\int_{1}^{\omega}
\frac{d\zeta}{\zeta^{2} \left(\zeta^3+5\sqrt{2}-\zeta^{-3}\right)^{2/3}}=0
\]
and similarly $A_{42}=0$. The two integrals $A_{12}$ and
$A_{32}$ can be expressed in terms of the hypergeometric function
$_{2}F_{1}$. Letting
$F:={}_{2}F_{1}(\frac{1}{6},\frac{2}{3};1;-\frac{2}{25})$,
we find
\[
A_{12}=-\frac{2i\bar{\omega}}{3(5\sqrt{2}\alpha)^{1/3}}
\int_{-\frac{\pi}{2}}^{\frac{\pi}{2}}\frac{du}{\left(1-\frac{i\sqrt{2}}{5}\sin
u\right)^{1/3}}=-\frac{2\pi i\bar{\omega}F}{3\sqrt[3]{5}\sqrt[6]{2}
\alpha^{1/3}}
\]
and, using the relation (see \cite{AS}, p. 559)
\[
_{2}F_{1}\left({\frac{1}{3}},{\frac{5}{6}};1;{-\frac{2}{25}}\right)=
\frac{\sqrt[3]{5}}{\sqrt{3}}{\,}
_{2}F_{1}\left({\frac{1}{6}},{\frac{2}{3}};1;{-\frac{2}{25}}\right),
\]
we obtain
\[
A_{32}=\frac{2i{\omega}}{3(5\sqrt{2}\alpha)^{2/3}}
\int_{-\frac{\pi}{2}}^{\frac{\pi}{2}}\frac{du}{\left(1-\frac{i\sqrt{2}}{5}\sin
u\right)^{2/3}}=\frac{2\pi i{\omega}F}{3 \sqrt{3}\sqrt[3]{10}\alpha^{2/3}}.
\]

Our choice of $a_{1}$ and $b_{1}$ implies that the
periods around these two cycles are related by conjugation,
\[
A_{i1}=\overline{B_{i1}}.
\]
This follows from the fact that the paths $\pi \circ  a_{1}$ and
$\pi \circ b_{1}$ are complex conjugate, while
\[ 
\overline{\eta_{2}(\zeta)}={\eta_{3}(\overline{\zeta})}
\]
from the definition in (\ref{etaj}).

There remain eight integrals to be calculated.
By resorting to numerical integration, we have established that they
are related to the periods around $a_{2}$ by simple numerical factors.
The conclusion is that the two blocks $\mathbf{A}$ and 
$\mathbf{B}$ of the period matrix can be written as
\[
\mathbf{A}=\begin{bmatrix}
\displaystyle{\frac{2\pi F}{3\sqrt{3}\sqrt[3]{5}\sqrt[6]{2}\alpha^{1/3}}} &
\displaystyle{-\frac{2\pi i\bar{\omega} F}{3\sqrt[3]{5}\sqrt[6]{2}\alpha^{1/3}}} &
\displaystyle{-\frac{2\pi i\omega F}{3\sqrt[3]{5}\sqrt[6]{2}\alpha^{1/3}}} &
\displaystyle{-\frac{2\pi F}{3\sqrt{3}\sqrt[3]{5}\sqrt[6]{2}\alpha^{1/3}}}  \\[15pt]
0&
0&
0&
\displaystyle{\frac{4\sqrt{2}\pi\bar{\omega} F}{9\sqrt[3]{10}\alpha^{2/3}}} \\[15pt]
\displaystyle{\frac{2 \pi F}{9\sqrt[3]{10}\alpha^{2/3}}} &
\displaystyle{\frac{2\pi i \omega F}{3\sqrt{3}\sqrt[3]{10}\alpha^{2/3}}} &
\displaystyle{\frac{2\pi i \bar{\omega}F}{3\sqrt{3}\sqrt[3]{10}\alpha^{2/3}}} &
\displaystyle{-\frac{2 \pi F}{9\sqrt[3]{10}\alpha^{2/3}}} \\[15pt]
\displaystyle{\frac{4\sqrt{2}\pi\omega F}{9\sqrt[3]{10}\alpha^{2/3}}} &
0&
0&
0
\end{bmatrix}
\]
and
\[
\mathbf{B}=\begin{bmatrix}
\displaystyle{\frac{2\pi  F}{3\sqrt{3}\sqrt[3]{5}\sqrt[6]{2}\alpha^{1/3}}} &
\displaystyle{-\frac{4\pi F}{3\sqrt{3}\sqrt[3]{5}\sqrt[6]{2}\alpha^{1/3}}} &
\displaystyle{-\frac{4\pi  F}{3\sqrt{3}\sqrt[3]{5}\sqrt[6]{2}\alpha^{1/3}}} &
\displaystyle{\frac{2\pi  F}{3\sqrt{3}\sqrt[3]{5}\sqrt[6]{2}\alpha^{1/3}}} \\[15pt]
0&
\displaystyle{\frac{4\sqrt{2}\pi i  F}{9\sqrt{3}\sqrt[3]{10}\alpha^{2/3}}} &
\displaystyle{-\frac{4\sqrt{2}\pi i F}{9\sqrt{3}\sqrt[3]{10}\alpha^{2/3}}} &
\displaystyle{-\frac{4\sqrt{2}\pi \omega F}{9\sqrt[3]{10}\alpha^{2/3}}} \\[15pt]
\displaystyle{\frac{2 \pi F}{9\sqrt[3]{10}\alpha^{2/3}}} &
\displaystyle{-\frac{4\pi i F}{9\sqrt{3}\sqrt[3]{10}\alpha^{2/3}}} &
\displaystyle{\frac{4\pi i  F}{9\sqrt{3}\sqrt[3]{10}\alpha^{2/3}}} &
\displaystyle{\frac{2 \pi  F}{9\sqrt[3]{10}\alpha^{2/3}}} \\[15pt]
\displaystyle{\frac{4\sqrt{2}\pi\bar{\omega} F}{9\sqrt[3]{10}\alpha^{2/3}}} &
\displaystyle{-\frac{4\sqrt{2}\pi i F}{9\sqrt{3}\sqrt[3]{10}\alpha^{2/3}}} &
\displaystyle{\frac{4\sqrt{2}\pi i F}{9\sqrt{3}\sqrt[3]{10}\alpha^{2/3}}} &
0
\end{bmatrix}.
\]

We have now all that is needed to determine $\alpha$ from the conditions
(\ref{explES}). For a given $\alpha$, this is a system of eight real
linear equations in eight (integer) unknowns. It has a solution given
by
\[
\mathbf{n}=\begin{bmatrix}0\\0\\0\\0\end{bmatrix},\;\;\;\;\;
\mathbf{m}=m\begin{bmatrix}0\\1\\1\\0\end{bmatrix},
\]
where $m$ satisfies
\[
m=\frac{3\sqrt{3}\sqrt[3]{5}\sqrt[6]{2}\,\alpha^{1/3}}{4 \pi F}.
\]
Now $m$ must be either $1$ or $-1$ for
$(\mathbf{m},\mathbf{n})$ to be primitive in $\mathbb{Z}^{\oplus 8}$.
If we take $m = 1$,
\begin{equation} \label{atetra}
\alpha= \frac{32\sqrt{2}\pi^{3}F^{3}}{405 \sqrt{3}},
\end{equation}
while $m=-1$ reverses the sign of $\alpha$. The two solutions can
be seen to
be a rotation of each other by $\zeta\mapsto -\frac{1}{\zeta}$;
to fix ideas, we take $\alpha$ positive from now on.
It can be checked numerically that (\ref{atetra}) agrees with the solution 
obtained in \cite{HMM}. For our orientation of the spectral curve 
(\ref{ch3ansatz}), the latter is given by
\begin{equation} \label{theira}
\alpha= \frac{\sqrt{2}}{3\sqrt{3}} \cdot
\frac{\Gamma(\frac{1}{6})^{3}\Gamma(\frac{1}{3})^{3}}
{48\sqrt{3}\pi^{3/2}}=
\frac{\Gamma(\frac{1}{3})^{9}}{48\sqrt{6} \pi^{3}}.
\end{equation}
In section \ref{4hedralact}, we use a change
of variables projecting $S$ onto an elliptic curve to relate
analytically the two results.

\begin{figure}[ht]
\begin{center}
\epsfig{file=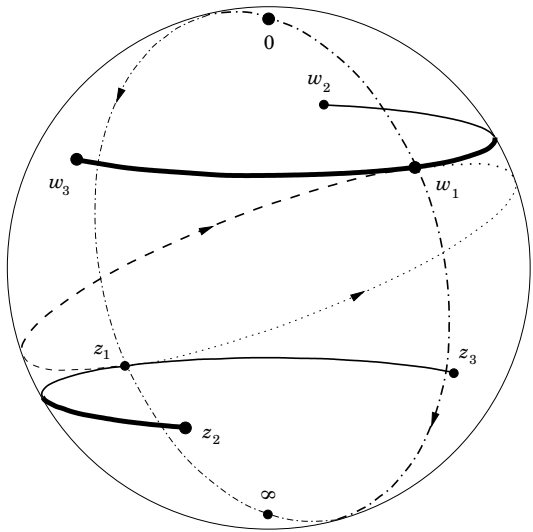,width=7cm}
\caption{{\small \it The special 1-cycle $c$ for the tetrahedral 3-monopole}}
\label{figcycle}
\end{center}
\end{figure}

The special 1-cycle $c$ for the tetrahedral 3-monopole with $a$
positive is 
\begin{equation} \label{invc}
c=b_{2}+b_{3}.
\end{equation}
It is sketched on the Riemann sphere in Figure~\ref{figcycle}, after
simplification using relations in homology --- for example, the sum of
the three equators with the same orientation is homologous to zero,
which is clear from Figure~\ref{figspech3}. 
Clearly, it is not a combination of a lift of equators, since the
projections of $b_{2}$ and $b_{3}$ enclose different branch points on
the Riemann sphere.
In the next section, it will be proved that $c$ is invariant under the
tetrahedral group.

\subsection{Action of the tetrahedral group} \label{4hedralact}

The spectral curve $S$ defined by (\ref{ch3ansatz}) admits an action
of the tetrahedral group $\mathfrak{A}_{4}\subset SO(3)$ determined by $PSU(2)$
transformations on $\zeta$; the corresponding rotations are the symmetries
of the tetrahedron drawn in Figure~\ref{figvrtxup}. This induces
an action on $H_{1}(S,\mathbb{Z})$, which we now describe.

Recall that $\mathfrak{A}_{4}$ is generated by the 3-cycle $(123)$
and the double transposition $(12)(34)$. We represent these as the rotation
by~$\frac{2\pi}{3}$ about the direction defined by the top vertex $0$
of the tetrahedron in Figure~\ref{figvrtxup},
\begin{equation} \label{genr}
R:\zeta \mapsto e^{\frac{2\pi i}{3}}\zeta,
\end{equation}
and the rotation by $\pi$ around the axis connecting the edge
midpoints $w_{1}$ and $z_{1}$,
\begin{equation} \label{gent}
T:\zeta \mapsto \frac{\sqrt{2}-\zeta}{1+\sqrt{2}\zeta},
\end{equation}
respectively. 
Later, we will also be interested in another element of order two,
\begin{equation} \label{genv}
V=R\,T R^{\,2},
\end{equation}
which corresponds to a rotation by $\pi$ about the axis connecting
$w_{2}$ and  $z_{2}$.
We also denote by $R$, $T$ and $V$ the maps induced on $S$ by
(\ref{genr}), (\ref{gent}) and (\ref{genv}).

On the complex plane, $R$ is of course just the rotation by~$\frac{2\pi}{3}$
about the origin, while $T$ and $V$ are elliptic M\"obius
transformations of order two with $w_{1}, z_{1}$ and 
$w_{2}, z_{2}$ as fixed points, respectively.
A way to visualize the action of $T$ or $V$ is to draw the
(invariant) circles of Apollonius corresponding to the two fixed 
points; the other four branch points of $\pi|_{S}$ all lie on one of
these circles and it is easy to verify that they are permuted as expected
under the two transformations.

\begin{figure}[ptbh]
\begin{center}
\epsfig{file=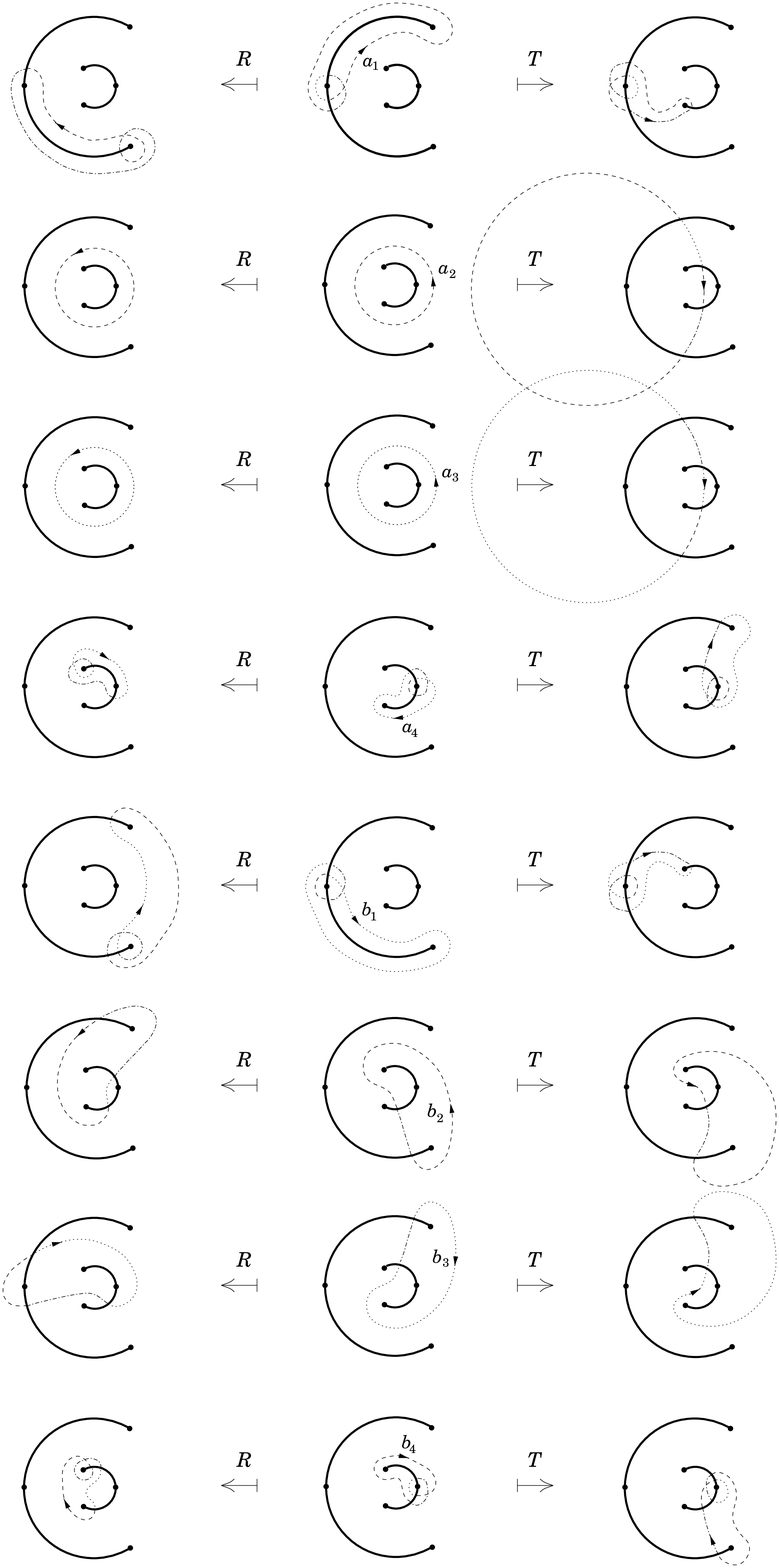,height=20cm}
\caption{{\small \it The action of $R$ and $T$ on the basis of 
$H_{1}(S,\mathbb{Z})$}}
\label{figmaprt}
\end{center}
\end{figure}

To describe the action of $\mathfrak{A}_{4}$ on $H_{1}(S,\mathbb{Z})$, we
start by computing the matrices representing the generators $R$ and
$T$. The effect of $R$ is easy to understand, since it leaves the three
annuli over $C=\{\zeta :|w_{1}|<|\zeta|<|z_{1}|\}$ invariant. $T$ is
harder to describe since it does not preserve the annuli, on which we can
easily keep track of the sheet labels by using the expression
(\ref{etaj}) for $\eta_{j}(\zeta)$. But we can still use (\ref{etaj}) when
$\zeta$ is in the smaller region $C_{+} \cup C_{-}$, where
\[
C_{\pm}:=\{\zeta \in C\cap T(C): \pm {\rm Im} \:\zeta >0\} 
\]
are mapped onto each other by $T$. Denoting by $C_{\pm,j}$
the intersection of $(\pi|_{S})^{-1}(C_{\pm})$ with sheet $j$, it can
be concluded that $T$ sends $C_{\pm,j}$ to $C_{\mp, j\mp 1}$, where the
labels are taken mod $3$. The sheet that contains the image
under $R$ or $T$ of any point on $S$ can now be easily identified from
these data and analytic continuation. In particular, we conclude that
the 1-cycles in our basis for $H_{1}(S,\mathbb{Z})$ are mapped as shown in
Figure~\ref{figmaprt}.

We can now use the perfect intersection
pairing (\ref{h1basis}) to compute the matrices of the maps $R_{*}$
and $T_{*}$ induced on homology from the intersection numbers of the 
1-cycles $a_{i}$ and $b_{i}$ with
their images. Let $c_{i}:=a_{i}$ and $c_{4+i}:=b_{i}$ for
$i=1,\ldots,4$.
Defining 
\[
M_{ij}:=\sharp (R_{*}c_{i},c_{j}),\;\;\;\;\; 
N_{ij}:=\sharp(T_{*}c_{i},c_{j}),
\]
we obtain the entries of the matrices  $\mathbf{R}$ and $\mathbf{T}$ 
representing $R_{*}$ and $T_{*}$ as
\[
R_{ij}=\sum_{k=1}^{8}J_{ik}M_{jk},\;\;\;\;
T_{ij}=\sum_{k=1}^{8}J_{ik}N_{jk},
\]
where, as in (\ref{jmatrix}),
\[
J_{ij}:=\sharp(c_{i},c_{j})
=\begin{bmatrix}
 & \mathbf{1}_{4} \\
-\mathbf{1}_{4} & 
\end{bmatrix}_{ij} .
\]
The intersection numbers $M_{ij}$ and $N_{ij}$ can be just
read off from Figure~\ref{figmaprt}, and we get
\[
\mathbf{R}=\left[
\begin{array}{cccc|cccc}
 0& 0& 0& 0&-1& 0& 0& 0\\
 0& 1& 0& 0&-1& 0& 0&-1\\
 0& 0& 1& 0& 1& 0& 0& 1\\
 0& 0& 0& 0& 0& 0& 0& 1\\ \hline
 1& 0& 0& 0&-1&-1& 1& 0\\
 0& 0& 0& 0& 0& 1& 0& 0\\
 0& 0& 0& 0& 0& 0& 1& 0\\
 0& 0& 0&-1& 0& 1&-1&-1
\end{array}
\right]
\]
and
\[
\mathbf{T}=\left[
\begin{array}{cccc|cccc}
-1& 0&-1& 0& 0& 0& 0& 0\\
 0& 1& 0& 0& 0& 0& 0& 0\\
 0& 0& 1& 0& 0& 0& 0& 0\\
 0&-1& 0&-1& 0& 0& 0& 0\\ \hline
 0& 1& 0& 0&-1& 0& 0& 0\\
-1& 0&-1& 0& 0& 1& 0&-1\\
 0& 1& 0& 1&-1& 0& 1& 0\\
 0& 0&-1& 0& 0& 0& 0&-1
\end{array}
\right].
\]
So the characters of the $\mathfrak{A}_{4}$ representation on 1-cycles are 
\[
{\rm tr}\,\mathbf{1}_{8}=8,\;\;\;\;
{\rm tr}\, \mathbf{R}=2= {\rm tr}\, \mathbf{R}^2,\;\;\;\;
{\rm tr}\,\mathbf{T}=0,
\]
and this shows that $H_{1}(S,\mathbb{C})$ splits as
$\mathbf{1}^{\oplus 2}\oplus \mathbf{3}^{\oplus{2}}$.
Another way to see this is to consider the action of $\mathfrak{A}_{4}$ by
pull-back on the holomorphic 1-forms 
$\Omega^{(\ell)}$ by $R$ and $T$ and calculate the characters to
conclude that $H_{0}(S,\Omega_{S}^{1})$ splits as 
$\mathbf{1}\oplus \mathbf{3}$ under $\mathfrak{A}_{4}$ (with
$\Omega^{(1)}$ spanning the
trivial singlet and being orthogonal to the triplet), and use Poincar\'e\
duality.

Using the matrices for $R_{*}$ and $T_{*}$, we can compute the
projection $\boldsymbol{\Pi}$ onto the subspace
$\mathbf{1}^{\oplus 2} \subset H_{1}(S,\mathbb{C})$ as
\[
\boldsymbol{\Pi} = 
\frac{1}{|\mathfrak{A}_{4}|}\sum_{\sigma \in \mathfrak{A}_{4}}
\boldsymbol{\sigma}= \frac{1}{4} \left[
\begin{array}{cccc|cccc}
  0& -1& -1&  0&  0&  0&  0&  0\\
  0&  2&  2&  0&  0&  0&  0&  0\\
  0&  2&  2&  0&  0&  0&  0&  0\\
  0& -1& -1&  0&  0&  0&  0&  0\\ \hline
  0&  1&  1&  0&  0&  0&  0&  0\\
 -1&  0& -3&  1& -1&  2&  2& -1\\
 -1&  3&  0&  1& -1&  2&  2& -1\\
  0& -1& -1&  0&  0&  0&  0&  0
\end{array}
\right].
\]
The range of this matrix is spanned by
\begin{equation} \label{invars}
\begin{bmatrix}0&0&0&0&0&1&1&0\end{bmatrix}^{t} 
\;\;\;\;\;{\rm and}\;\;\;\;\;  
\begin{bmatrix}2&-4&-4&2&-2&3&-3&2\end{bmatrix}^{t},
\end{equation}
so we conclude that the special cycle $c$ given in (\ref{invc}) is 
invariant under the action of $\mathfrak{A}_{4}$. Notice that in
(\ref{invars}) the first vector is antisymmetric whereas the second is
symmetric under reality.

We can explore the action of the Vierergruppe $D_{2} \subset \mathfrak{A}_{4}$ 
generated by the two elements $T$ and $V$ to express the value
$\alpha$ given by (\ref{atetra}) in terms of elliptic integrals, as
in \cite{HMM}. 
The actions of both $T$ and $V$ are much easier to describe in an
alternative orientation of the monopole, obtained
by rotation of (\ref{ch3ansatz}) under 
\[
\zeta \mapsto \frac{(z_{3}-z_{1})(\zeta -w_{1})}
{(z_{3}-w_{1})(\zeta-z_{1})}.
\]
Then the spectral curve is taken to the form
\begin{equation} \label{edgeup}
\eta^{3}+\frac{3\sqrt{3}}{\sqrt{2}}\alpha i\zeta(\zeta^{4}-1)=0,
\end{equation}
which can be described as a covering of $\mathbb{CP}^{1}$ with
branch points at $0$, $\pm 1$, $\pm i$ and $\infty$. In this
configuration, $T$ is just $\zeta \mapsto -\zeta$, while $V$ 
is $\zeta \mapsto -\frac{1}{\zeta}$. The map
\[
p: \zeta \mapsto \frac{1}{2} \left( \zeta^{2} + \frac{1}{\zeta^{2}}
\right) =:z
\]
identifies points in the same orbit of $D_{2}$, having the first
quadrant as fundamental region. Under the map induced on
$T'\mathbb{CP}^{1}$ by $p$, the
spectral curve 
(\ref{edgeup}) goes to
\[
w^3 + 24\sqrt{6}\alpha i(z^{2}-1)^2=0,
\]
which is a torus by the Riemann--Hurwitz formula and corresponds to
the quotient $S/D_{2}$. The two pairs of branch cuts on the original
Riemann sphere are both identified with a cut connecting the new branch
points $1$, $\infty$ and $-1$ along the real axis.
With some care, it can be seen that the image of the 1-cycle $c$ in 
(\ref{invc}) can be identified with a cycle going four times along the
imaginary axis in the negative direction, on the sheet
containing the point $(w,z)=(2\sqrt[6]{2}\sqrt{3}\alpha^{1/3} i,0)$. On the
other hand, it is easy to see that the 1-form $\Omega^{(1)}$ is
given by the same expression in the new orientation, and
\[
p^{*}\left( \frac{dz}{3w}\right)=\frac{d\zeta}{3\eta}=\Omega^{(1)}.
\]
Thus we can write
\[
\oint_{c}\Omega^{(1)}=\oint_{p_{*}c}\frac{dz}{3w}=
4\int_{i \infty}^{-i \infty}\frac{dz}
{6 \sqrt[6]{2}\sqrt{3}\omega \alpha^{1/3} \left(-i(z^{2}-1)^{2}\right)^{1/3}}
\]
and this can be reduced to an elliptic integral, yielding
\[
-\frac{\Gamma(\frac{1}{3})^{3}}
{\sqrt{6} \pi \alpha^{1/3}}.
\]
Now this has to be equal to $-2$ by (\ref{justleq1}). 
Thus we get
\[
\alpha=\frac{\Gamma(\frac{1}{3})^{9}}
{48\sqrt{6} \pi^{3}}
\]
in agreement with (\ref{theira}).

\section{Discussion} \label{cnclsn}

The version of the Ercolani--Sinha constraints that we derived in 
section~\ref{ErcSin} generalises the Corrigan--Goddard conditions to
all monopoles with a nonsingular spectral curve. 
An interesting aspect is the existence of a distinguished 1-cycle $c$
on the spectral curve.
The premises in the 
Corrigan--Goddard approach lead to the constraint (\ref{CGc}) for
$c$, but their conditions
are otherwise equivalent to 
equation (\ref{ourES}). In section \ref{4hedral}, we have applied
(\ref{ourES}) to rederive the scale parameter $\alpha$ in
the spectral curve of the tetrahedrally symmetric monopole of charge 3. 
We also verified that this monopole provides an example where
our condition (\ref{ourES}) can be satisfied but those of
Corrigan and Goddard are not.

Let us make some remarks about the nature of the special 1-cycle $c$.
Given a nonsingular spectral curve $S$ in $\mathbb{T}$, $c$ is uniquely
determined as the solution to equation (\ref{ourES}); we have
established that it is always antisymmetric under the real structure.
Moreover, although the left-hand side of (\ref{ourES}) depends on the spatial 
orientation of the monopole, $c$ remains constant along the $SO(3)$ orbit 
of $S$ in the moduli space ${\cal N}_{k}$. 
In fact, its components in a given homology basis
are integer solutions to a linear equation and cannot change when the 
spectral curve is rotated, since the period matrix occurring in
(\ref{quant1}) never becomes singular.
This argument applies to more general deformations in ${\cal N}_{k}$
that do not pass through monopoles with a singular spectral curve.
It also implies that $c$  has to be invariant under any
rotational symmetry of the spectral curve, and this imposes further 
restrictions --- for example, in the case of the tetrahedrally
symmetric 3-monopole that we studied in section \ref{4hedral}, 
this consideration together with the $\tau$-antisymmetry completely
determines $c$ up to sign.

As implied in \cite{ES}, the components of the 1-cycle $c$ are the
characteristics of the line bundle $L^{2}|_{S}$ and can thus be interpreted
as giving the direction of the linear flow determined by Nahm's equations on
the Jacobian of the spectral curve $S$. 
Another interpretation for $c$ is afforded by equation (\ref{defmn}).
Recall that the triviality of $L^{2}|_{S}$
provides for two nowhere vanishing functions $f_{0}$
and $f_{1}$ on the open sets $U_{0}\cap S$ and $U_{1}\cap S$. We may
wonder whether we can define logarithms of these functions. And of
course the answer is no: the nonzero components of $c$ correspond to
nontrivial periods of both $d\, {\rm log} f_{0}$ and $d\, {\rm log}
f_{1}$, and so they cannot be exact 1-forms. To define the logarithms,
one should eliminate the 1-cycles correponding to the nonzero
periods, by cutting $S$ along their conjugate homology 1-cyles in 
the canonical basis (\ref{h1basis}). But we can see from (\ref{ESc}) 
that this is equivalent to cutting $S$ along $c$. The Riemann surface
of ${\rm log}f_{0}$ or ${\rm log}f_{1}$ is then obtained from the cut
surfaces $U_{0}\cap S$ or $U_{1}\cap S$ by analytic continuation across the
cuts, and this yields an infinite cover of the original open sets. 
So we may regard $c$ as a topological obstruction to defining the 
logarithms of the nowhere vanishing functions $f_{0}$ and $f_{1}$ on
the spectral curve punctured at the points lying over $\zeta=\infty$
and $\zeta=0$, respectively.

We should emphasise that the Ercolani--Sinha algorithm is still not
sufficient to ensure smoothness of the fields if $k>2$, since it does not 
include the condition (\ref{smooth}). The family of nonsingular
spectral curves of monopoles has codimension zero in the family of
real curves in $|{\cal O}(2k)|$ satisfying
equation (\ref{ourES}), 
but the inclusion is proper in general. For example, it
can be shown that the icosahedrally symmetric curve
\begin{equation}\label{pseudo6}
\eta^{6} + \alpha \zeta (\zeta^{10}+11\zeta^{5}-1)=0
\end{equation}
satisfies (\ref{ourES}) for some constant $\alpha$, but not
(\ref{smooth}); this follows from the conclusion in \cite{HMM} that
there is no 6-monopole with icosahedral symmetry. It is known \cite{HSico}
that an icosahedrally symmetric monopole of charge $7$ exists, and its
spectral curve is reducible to a projective line and a smooth genus $25$ 
curve of the form (\ref{pseudo6}), with $\alpha=\frac{3^{3} 
\Gamma(\frac{1}{3})^{18}}{2^{8}\pi^{6}}$.

An interesting question is to understand how (\ref{ourES}) 
degenerates when a spectral curve 
becomes singular. Some 
singularities arise by imposing interesting symmetries on the
monopoles, as in the case of the axially symmetric monopoles that
we have mentioned already. We may expect that the condition still 
holds for other singular spectral curves, but it is not clear how the 
1-cycle $c$ is to be determined in general. 

\vspace{3cm}
\noindent
{\large \bf{Acknowledgements}}
\vspace{.5cm}

\noindent

We thank Roger Bielawski for advice. 
CJH thanks Fitzwilliam College, Cambridge, 
for a research fellowship.
NMR is supported by Funda\c{c}\~ao para a Ci\^encia e a
Tecnologia, Portugal, through the research grant BD/15939/98.

\end{document}